\documentclass[%
 aip,
 jmp,%
 amsmath,amssymb,
 reprint,%
]{revtex4-1}

\usepackage{graphicx}% Include figure files
\usepackage{dcolumn}% Align table columns on decimal point
\usepackage{bm}% bold math
\begin{document}

\title{Acoustic analog of Hall effect in superconductive films}
\author{ E. D. Gutliansky }
 \altaffiliation {Physics Institute of Southern Federal University,  Rostov-on-Don, 344090, Russia}
 \email{gutlian@yandex.ru}
%\thanks{}
\affiliation{}
\date{\today}

\begin{abstract}
Longitudinal electric field of a surface acoustic wave (SAW) drags vortex structure of a superconductive film,
deposited on a piezoelectric substrate, and generates longitudinal DC component of an acoustoelectric field,
which does not depend on direction of an external magnetic field. The contra-directional vortices are dragged by SAW in opposite directions.
This phenomenon represents an acoustic analog of Hall effect, where vortices are an analog of current carriers, and the SAW Pointing vector acts
as an impressed electric field. The calculation of the acoustoelectric field for a YBCO film with niobate lithium substrate coincides well
with experimental data.
\end{abstract}

\pacs{}% insert suggested PACS numbers in braces on next line

\maketitle

Ultrasonic waves (UW), spreading in superconductors with dynamic
vortex structure (VS), drag it and induce constant component of an
electric field. This effect was observed during the experiments
[\cite{ilya}, \cite{il-le}, \cite{zav}, \cite{huch}, \cite{huJa}],
and was researched in theory [\cite{gut94}, \cite{gut98},
\cite{gut01}]. We name it an acoustoelectric effect on the analogy
of electron-phonon drag effect in semiconductors and metals
[\cite{par}].

This effect is interesting enough from the experimental point of
view. The effect makes it possible to introduce new measuring ways
of VS parameters. Namely, we are able to measure VS density
derivative and VS viscosity coefficient. There is another reason
for our interest. The effect gives us a way of controlling
vortices movement by means of ultrasonic waves. It enables us to
create a new class of devices using back reaction effects
(changing SAW parameters at the expense of its' interaction with
moving VS [\cite{gut02}, \cite{gut05}, \cite{gut07},
\cite{gutst}]. We may call this science area 'acoustovihronics' by
analogy with 'acoustoelectronics' [\cite{gul}].

Acoustoelectric effect in YBCO films with niobate lithium
substrate was observed in the works [\cite{ilya} - \cite{il-le}].
There was a standard acoustoelectric effect above the transition
point in superconductive state: entrainment of normal carriers by
UW. Acoustoelectric field sign and Hall effect sign proved that
there was a hole conductivity type.  But the authors discovered a
sign reversal (disappearance of effect was expected) in lower
temperatures during the transition through the transition point in
superconductive transition.

A theory able to present qualitative explanation of the experiment
[\cite{ilya} - \cite{il-le}] was proposed in [\cite{gut98}]. The
appearance of longitudinal electric field was explained as
transversal movement of vortex structure, induced by UW.  The
effect's value was uniquely determined with $\alpha$ parameter
which is proportionate to Hall coefficient in superconductors with
VS. In this theory the factor acting on VS was UW deformation
vector. The theory gave right qualitative description of the
effect.   Nontheless, the usability of the theory for explanation
of the discussed experiment is doubtful, because it is difficult
to consider the films which were used in the experiment to be
'clean', namely, demonstrating Hall effect in vortex phase.  The
fact that longitudinal SAW is accompanied by longitudinal electric
field was not taken into account [\cite{gut98}]. In a
superconductor at least a longitudinal alternating electric field
is known to penetrate on a depth $\lambda _L $  ($\lambda _L $ -
London penetration depth), and we may consider that the field
coincides with SAW field in the films with $d \le \lambda _L $
thickness. This field will induce longitudinal supersonductive
currents which create a new interaction mechanism between SAW and
vortices.  This mechanism was not discussed in [\cite{gut98}].
The mechanism must work always, regardless of Hall effect in a
superconductive film. Therefore it is reasonably to believe, that
the interaction would result in an entrainment of VS in the
transversal direction and appearance of a longitudinal electric
field in the 'dirty' films.

 Moreover, SAW electric field creates another interesting possibility of contrillong VS.
 It separates vortices of opposite orientation, since generated by SAW superconductive currents exert forces of opposite directions
 to vortices opposite orientations. SAW drags different oriented vortices along the line perpendicular to SAW propagating in the opposite directions
 (see Fig.1),  some of them accumulate on the one side of the substrate, others – on the other.  Above described effect is similar to Hall affect,
 where a flux of SAW energy acts on opposite directed vortices as electric current on carriers with different signs.

The goals of the present work are:

1. demonstrate, that SAW electric field drags  superconductive
film VS in transversal direction and its' movements, oscillations
in this direction generate constant component of electric field in
longitudinal direction (longitudinal acoustoelectric effect);

2. prove, that the effect must be observed in 'dirty'
superconductive films with piezoelectric substrate.

3. make certain, that taking into account the electric field of
SAW explains the experiment [\cite{ilya} - \cite{il-le}] both
qualitatively and quantatively without assumption existence of
Hall effect of VS in superconductive film. All physical effects
named below are novel and have not been discussed in scientific
literature before.

We adopt the following picture of current carriers and ion motion
in the film induced by SAW (current curriers can be electrons or
holes). The current carriers liquid consists of two parts: normal
and superconductive. The normal component is moved together with
film ions dragged   by viscous forces responsible for the normal
resistance. In this picture the role of normal carriers is reduced
only to partial screening of ions.  Condition of applicability of
this model is given by an inequality $\omega \tau  \ll 1$ ($\tau
$, $\omega $ are relaxation time of the normal current carriers,
SAW frequency, correspondingly). We describe motion of film VS
within the framework of hydrodynamic approach which is valid for
SAW wave lengths that are much larger than intervortex distance,
so that the  VS deformation vector $\vec W$  can be regarded as
continuous function of coordinates.

We will consider the following experimental geometry [\cite{ilya},
\cite{il-le}]. Let SAW propagates along positive direction of Z
axis of the substrate of Y – niobate lithium cut. External
magnetic field $\vec B_0 $  is directed along positive direction
of X axis. The surface is covered with superconductive film with
thickness $d \le \lambda _L $  and $d,\lambda _L  \ll \lambda _R $
($\lambda _R $ is SAW wave length). This condition enables us to
assume SAW electric field within the superconductive film $\vec
E_{ext} $ homogeneous and we assume it to be equal with the field
on the surface of the substrate $\vec E_{ext}  = \vec E_{ext0}
\exp \left( {ikz - i\omega t} \right)$.
 Deformation vector of ion lattice within the film is equal to substrate surface deformation vector
 $\vec U = \vec U_0 exp\left( {ikz - i\omega t} \right)$,
  where k  is SAW wave vector.

London equation for superconductive $\vec J_S $  current generated
by field $\vec E_{ext} $
 and VS movement in the laboratory coordinate system is:
\begin{equation}
\lambda _L^2 \mu _0 {{\partial \vec J_S } \over {\partial t}} =
\vec E + \vec E_{ext}  - \vec B_v  \times \dot {\vec W}
\label{eq:one}
\end{equation}
where  $\vec E$  - internal electric field, $\vec B_v $ - magnetic
induction produced by the VS of the film, $\vec W$  - VS
deformation vector, $\vec B_v  = \phi _0 n_\nu  \vec e$, $n_v $ -
2D vortex density: number of vortices per square unit in the
plane, perpendicular to vortex lines, direction $\vec B_v $  is
determined by vector $\vec e$, tangent to vortex line

Applying the operator $\nabla  \times $  to both sides of
Eq.~(\ref{eq:one}), and using Maxwell's equation
\begin{equation}
\nabla  \times \vec E =  - {{\partial \vec {\rm B}} \over {\partial t}}
\label{eq:two}
\end{equation}
we obtain:
\begin{equation}
\lambda _L^2 \mu _0 \nabla  \times {{\partial \vec j_s } \over {\partial t}} =
  - {{\partial \vec B } \over {\partial t}} + \nabla  \times \vec E_{ext}  - \nabla  \times \left( {\vec B} _\nu
   \times \dot {\vec W} \right)
\label{eq:three}
\end{equation}
Applying the operator $\nabla  \times $  to  Eq.~(\ref{eq:three}),
and taking into account connection between total current in
superconductor and the ionic current  with superconductive current
[\cite{gut94}]
\begin{equation}
\vec J_s  = \vec J + qn_s \dot {\vec U} \label{eq:four}
\end{equation}
after simple manipulations we obtain the following expression:
\[{\partial  \over {\partial t}}\left(
{\lambda _L^2 \nabla ^2\vec{ F}_s  - \vec {j}_s}
 \right) =
 - qn_s \ddot {\vec U} - {1 \over
{\mu _0 }}\nabla  \times \nabla \times \vec E_{ext}+\]
\begin{equation}
{1 \over {\mu _0 }}\nabla  \times \nabla \times \left( {\vec B}
_\nu \times \dot {\vec W} \right)
\label{eq:five}
\end{equation}
Where $\mu _0 $, $n_s $, m, q  are magnetic susceptibility,
superconductive carriers density, their mass and charge. Let us
now write the local equation of motion of the VS (neglecting the
inertial mass of the vortices). Within the framework of
hydrodynamic approximation this equation follows from the
condition of balance of two forces:
\begin{equation}
\vec f_{fr}  = \vec f_M
\label{eq:six}
\end{equation}
Where $f_{fr}$, $f_m $  are the friction force of the VS against
the crystal lattice of the film and Magnus force densities,
respectively. The latter force has a hydrodynamic nature. It is
the result of interaction of superconductive currents with the
moving vortices and we name it Magnus force, though it is
necessary to note, that there are different interpretations of
this term, see for example [\cite{gut05}]. If VS is stationary
this force is called Lorenz force. $f_{fr} $  is the function of
the VS local velocity relative to superconductor
 $(\dot {\vec W} - \dot {\vec U})$  and  $(\dot {\vec W} - \dot {\vec U}) \times \vec B_v $.
 Expanding of $f_{fr} $  in these terms with accuracy to the second order, we can write it in the form
\begin{equation}
\vec f_{fr}  = \eta \left( {\dot {\vec W} - \dot {\vec U}} \right)
- \tilde \eta \left( {\dot {\vec W} - \dot {\vec U}} \right)
\times \vec B_v \label{eq:seven}
\end{equation}
where $\eta $  and $\tilde \eta $  are longitudinal and
transversal  VS viscosity, respectively.

Magnus force is equal to $\vec f_m  = \left( {\vec J_s  - qn_s
\dot {\vec W}} \right) \times \vec B_v $.

It is the result of Lorenz force expression $\vec f_L  = \vec J'_s
\times \vec B_v $  in local coordinate system, connected with
moving VS where $\vec J'_s $  is superconductive current in this
system. Placing (\ref{eq:seven}) and expression of $f_m $  to
(\ref{eq:six}) one obtains equation of VS
\begin{equation}
\eta \left( {\dot {\vec W} - \dot {\vec U}} \right) - \tilde \eta
\left( {\dot {\vec W} - \dot {\vec U}} \right) \times \vec B_v  =
\left( {\vec J_s  - qn_s \dot {\vec W}} \right) \times \vec B_v
\label{eq:eight}
\end{equation}
It is convenient to represent (\ref{eq:eight}) replacing $J_S $ on
the total current  $J$ (4) as
\begin{equation}
\eta \left( {\dot {\vec W} - \dot {\vec U}} \right) - \alpha
\left( {\dot {\vec W} - \dot {\vec U}} \right) \times \vec B_v  =
\vec J \times \vec B_v
\label{eq:nine}
\end{equation}
where  $\alpha  = qn_s  - \tilde \eta $.

In order to understand the physical meaning of coefficients $\eta $
 and $\alpha $, let us assume $\dot \vec U = 0$  and $J = const$  (no ultrasonic wave) and find conductivity $\sigma _\Omega  $
 and Hall coefficient $\sigma _H $  of superconductor in a mixed state from (\ref{eq:nine}).

Multiplying (\ref{eq:nine}) vectorial by $\vec B_v $  and taking
into account that $\vec E = \vec B_v  \times \dot {\vec W}$ -
electric field, induced by VS movement, one obtains:
\begin{equation}
\vec J_s  = {\eta  \over {B_v^2 }}\vec E - {\alpha  \over {B_v }}\vec e \times \vec E
\label{eq:ten}
\end{equation}
Consequently, specific conductivity is expressed through the
longitudinal viscosity $\sigma _\Omega   = {\eta  \over {B_v^2 }}$
and Hall coefficient is uniquely determinated by  $\alpha$
coefficient: $\sigma _H  = {\alpha  \over {B_v^{} }}$.

In "dirty" superconductors Hall effect is negligibly small and
therefore we can take $\alpha$ equal zero and hence  VS motion
equation will have the form
\begin{equation}
\eta \left( {\dot {\vec W} - \dot {\vec U}} \right) = \vec J
\times \vec B_v
\label{eq:eleven}
\end{equation}
Moreover, in order to solve the problem of VS  entrainment  by SAW
one needs continuity equation for VS, Maxwell equation and a
elasticity equation with term describing  coupling between the
ionic lattice deformation and  the VS motion
\begin{equation}
{{\partial \vec B _v } \over {\partial t}} = \nabla  \times \left(
\dot {\vec W} \times \vec B_v  \right) \label{eq:twelve}
\end{equation}
\begin{equation}
\nabla  \times \vec B = \mu _0 \vec J
\label{eq: thirteen }
\end{equation}
The elasticity equation was deduced in [\cite{gut98}].  In this
work we neglect back influence of VS on SAW and will assume, that
the deformation vector and the electric field of SAW are
predetermined.  We solve system of equations (\ref{eq:four},
\ref{eq:five}, \ref{eq:eleven} and \ref{eq:twelve}) with given SAW
parameters using step-by-step approach method assumed that $\vec
B_v  = \vec B_0  + \vec b_v $, $\dot {\vec W} = \dot {\vec W}_1 +
\dot {\vec W}_2 $  and find $\dot {\vec W}$.  Here $\vec B_0 $ is
a homogeneous component of superconductor magnetization equal to
external field, $B_0 $ is a projection of external field on axis
X, $\vec b_v $   is VS density fluctuations produced by SAW,
$\left| {\vec b_v } \right| \ll B_0 $, $\dot {\vec W}_1 $ , $\dot
{\vec W}_2 $  are the local velocities  of VS  in the first and in
the second orders of the SAW amplitude, respectively.

The SAW period time averaging of $\dot {\vec W}$ shows that in the
second order of the SAW amplitude, VS  has a constant velocity
both in SAW propagating  direction and in the perpendicular
direction.
\[
\left\langle {\dot {\vec W}_2 } \right\rangle _x  =  - {1 \over
2}k^2 \left( {1 - {{\eta _{,B} \left| {B_0 } \right|} \over {\eta
_0 }}} \right){1 \over {B_0 }}{{X^2 } \over {1 + X^2 }}U_0 \Phi
_0+
\]
\begin{equation}
 {1 \over 2}k\omega \left( {1 - {{\eta _{,B} \left| {B_0 }
\right|} \over {\eta _0 }}} \right){{m\omega } \over {qB_0 }}{{X^3
} \over {1 + X^2 }}U_0^2
 \label{eq:fourteen}
\end{equation}
\begin{equation}
\left\langle {\dot {\vec W}_2 } \right\rangle _z  = {1 \over
2}k\omega \left( {1 - {{\eta _{,B} \left| {B_0 } \right|} \over
{\eta _0 }}} \right){{X^2 } \over {1 + X^2 }}U_0^2
\label{eq:fifteen}
\end{equation}
where the brackets $\left\langle  \ldots  \right\rangle $   is an
averaging over SAW period, $\vec E_{ext}  =  - \nabla \Phi $,
$\Phi  = \Phi _0 \exp \left( {ikz - i\omega t} \right)$   is an
electric field potential, attended to SAW, $D = {{B_0^2 }
\mathord{\left/ {\vphantom {{B_0^2 } {\mu _0 }}} \right.
 \kern-\nulldelimiterspace} {\mu _0 }}\left( {1 + \lambda _L^2 k^2 } \right)$,
 $X = {{Dk^2 } \mathord{\left/ {\vphantom {{Dk^2 } {\omega \eta _0 }}} \right. \kern-\nulldelimiterspace} {\omega \eta _0 }}$,
 $\eta {}_0 = \eta \left( {\left| {B_0 } \right|} \right)$ is a zero order in VS viscosity coefficient expansion in fluctuations
 of its' density $b_v $,  $\eta _{,B}  = \left( {d\eta {{\left( {B_v } \right)} \mathord{\left/ {\vphantom {{\left( {B_v } \right)} {dB_v }}} \right.
 \kern-\nulldelimiterspace} {dB_v }}} \right)_{B_0 } $.

The first term in expression (\ref{eq:fifteen}) for the
transversal VS velocity appears due to SAW electric field. In case
similar experiment is conducted with the films with
non-piezoelectric substrate this term equals zero, but transversal
entrainment effect still exists, with other sigh. This effect is
marked with the second term (\ref{eq:fifteen}).  Its' physical
reason is connected with oscillations of the superconductor's
ionic lattice, which induce screen superconductive electronic
currents, which compensate magnetic field, induced by these
oscillations. These currents create Lorenz force which moves the
VS in the transversal direction. This mechanism works in bulk
superconductors as wall. Therefore a longitudinal ultrasonic wave
entrains VS in the transversal direction in bulk superconductors
and this effect is described by second term of
(\ref{eq:fourteen}).

The mechanisms discussed above determinate the VS entrainment
direction uniquely . This direction is specified by the sign
(\ref{eq:fourteen}). Separation of vortices resembles that in the
Hall effect – for example, vortices of one orientation deviate to
the right and another to the left. The use of this effect allows
separating vortices of different orientation – one type of
vortices will accumulate on the right side of the film, and
another – on the left. For example, in very thin films the
vortices of opposite orientation, arising as the result of
Kosterlitz Taules transition, can be separated by SAW. Originally
not magnetized film under the influence of SAW will get the
magnetisation of opposite signs on the its edges.

Obtained results show that SAW wave drags VS both in longitudinal
~(\ref{eq:fifteen}) and transversal directions
~(\ref{eq:fourteen}). Thus, longitudinal UW drags VS angularly to
the direction of its' spreading. For the first time expression
~(\ref{eq:fourteen}) was obtained in [\cite{gutdis}].

It is necessary to underline another peculiarity of dragging
effect in 'dirty' superconductors, namely: in case of Flux Flow
mode, if $\eta$ is proportional to B, then the SAW and
longitudinal UW in bulk superconductors does not drag at all,
since  ${{\eta _{,B} \left| {B_0 } \right|} \mathord{\left/
 {\vphantom {{\eta _{,B} \left| {B_0 } \right|} \eta }} \right.  \kern-\nulldelimiterspace} \eta }_0  = 1$.

The main observing quantity in the experiments with SAW
influencing on superconductors VS is the electric field, generated
by its' movement. The expression for this field, induced by VS
movement, is
\begin{equation}
\vec E = \vec B_v  \times \left( {\dot {\vec W} - \dot {\vec U}}
\right) \label{eq:sixteen}
\end{equation}
Now, substituting in (\ref{eq:sixteen}) the solutions of the
equation system
(\ref{eq:four},~\ref{eq:five},~\ref{eq:eleven},~\ref{eq:twelve})
and leaving in the obtained expression the terms only the first
and the second order infinitesimal on the SAW amplitude and
averaging this expression on the wave period,  we obtain a
constant component of acoustoelectric fields
\begin{equation}
\left\langle {\vec E} \right\rangle _x  = {1 \over 2}k\omega \left( {2 - {{\eta _{,B} \left| {B_0 } \right|} \over {\eta _0 }}} \right){{X^2 } \over {1 + X^2 }}U_0^2
\label{eq:seventeen}
\end{equation}
\begin{eqnarray}
\left\langle {\vec E} \right\rangle _z  =  - {1 \over 2}k^2 \left( {1 - {{\eta _{,B} \left| {B_0 } \right|}
\over {\eta _0 }}} \right){{X^2 } \over {1 + X^2 }}U_0 \Phi _0  +\nonumber \\
{1 \over 2}k\omega {{m\omega } \over q}\left( {2 - {{\eta _{,B}
\left| {B_0 }
 \right|} \over {\eta _0 }}} \right){{X^3 } \over {1 + X^2 }}U_0^2
\label{eq:eighteen}
\end{eqnarray}
Our first priority in this work is to find the constant
longitudinal acoustoelectric field in the direction of axis Z
which was observed in [\cite{ilya} - \cite{il-le}], but we also
provide expression for the transversal electric field
(\ref{eq:seventeen}). It is worth noting, that the latter was
observed in YBCO films in [\cite{huch}] and [\cite{huJa}].

The longitudinal acoustoelectric field (\ref{eq:eighteen}) is
independent from the sign of the external magnetic field. This
peculiarity allows to measure absolute number of vortices in
superconductor regardless of their orientation. Using tables
\cite{auld} we are able to express $U_0 $  and $\Phi _0 $  with
SAW power  per film width unit  $P_R $.
\begin{equation}
\left\langle {\vec E} \right\rangle _z  =  - {1 \over 2}k^2 \left( {1 - {{\eta _{,B} \left| {B_0 } \right|}
\over {\eta _0 }}} \right){{X^2 } \over {1 + X^2 }}{{P_R } \over \omega }23.75 \cdot 10^{ - 5}
\label{eq:nineteen}
\end{equation}
Numerical coefficients in the formula (\ref{eq:nineteen})
correspond with YZ niobate lithium section. In the formula
(\ref{eq:nineteen}) we neglect the term proportionate to
${{m\omega } \mathord{\left/ {\vphantom {{m\omega } q}} \right.
\kern-\nulldelimiterspace} q}$, since the working frequency in the
experiments [\cite{ilya} - \cite{il-le}] was 87 Mhz and it is far
less that the first term. The expression (19) differs only by the
numerical factor for different materials and cuts. All quantity
entering into it is presented in SI but the resulting electric
field is measured in ${{\mu V} \mathord{\left/ {\vphantom {{\mu V}
{cm}}} \right. \kern-\nulldelimiterspace} {cm}}$.

Fig.2 depicts the result of the calculations of longitudinal
acousoelectric effect for  Elisavsky and et. Experiment
[\cite{ilya}].

Solid line is a theory, squares are the experiment [\cite{ilya}].
In experiment [\cite{ilya}] film width is 1.8 mm, radiation power
- 2 watt, thus $P_R  = 1.41 \cdot 10^3 {w \mathord{\left/
{\vphantom {w m}} \right. \kern-\nulldelimiterspace} m}$.
Frequency = 87 Mhz, field = 1Ò, we used Tinkham ansatz for the
film specific resistance [13] $r = r_0 I_0^{ - 2} \left( {\gamma
_0 /2} \right)$, ${{\gamma _0 } \mathord{\left/ {\vphantom
{{\gamma _0 } 2}} \right. \kern-\nulldelimiterspace} 2} = \delta
\left( {1 - t} \right)^{3/2} B^{ - 1} $, $t = T/T_c $, $I_0 $ -
modified Bessel function, T - absolute temperature,
 $T_C $  - temperature of transition in  superconductive state, $r_0  = 1.6 \cdot 10^{ - 6} \Omega m$, viscosity coefficient was expressed
 with specific resistance: $\eta  = {{B^2 } \mathord{\left/ {\vphantom {{B^2 } r}} \right.  \kern-\nulldelimiterspace} r}$.
 Coefficient $\delta$ was used as an adjustable parameter and was equal to 57.

As follows from equation (\ref{eq:nineteen}) the sigh of the
longitudinal acoustoelectric field is negative. Above the
superconductive transition point this effect exists as well,
though it has positive sign because in this case vortices are
absent and the current carriers are normal holes. They entrain by
SAW and the relevant acoustoelctric field is positive. It was
observed in experiment [\cite{ilya}].

Thus we demonstrated that the SAW electric field drags vortices in
transversal direction and generates constant component of
longitudinal electric field and this field does not depend on the
direction of the external magnetic field. This peculiarity of the
effect provides us with the opportunity to measure a  whole number
of vortices in a film regardless of their orientation. The
obtained result allows us to explain the experimental data
\cite{ilya} both quantitatively and qualitatively. The theory, as
shown in the fig.2, describes the experiment perfectly.

\begin{figure}
\includegraphics{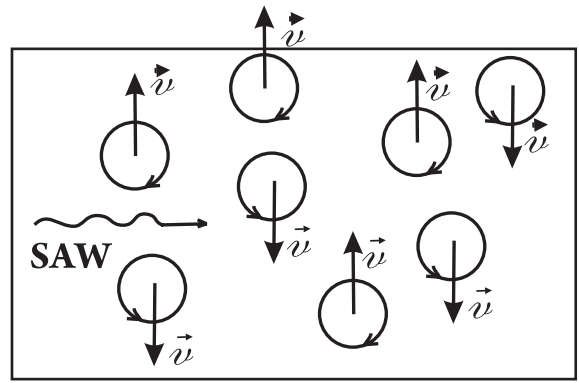}
\caption{\label{ris1} SAW electric field $\vec E_{ext} $ generates
superconductive current, which creates Lorenz force, which in its'
turn moves vortices of counterdirectional orientation in counter
directions. $V$–vortex velocity. Rings represent vortices, points
represent vortex orientation.}
\end{figure}

\begin{figure}
\includegraphics{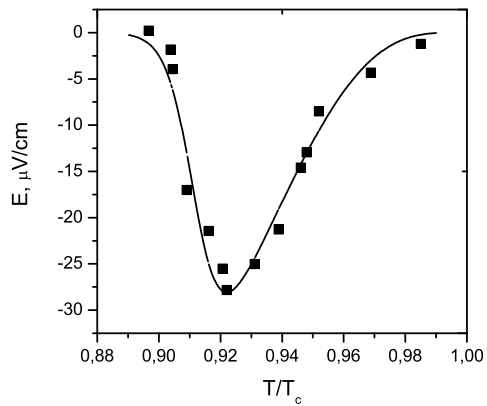}
\caption{\label{ris2} Longitudinal dependence of acoustoelectric
field on temperature. Temperature is presented with  relative
units $t = T/T_c $, T – temperature, $T_C $  - temperature of
transition in superconductive condition. Solid line is a
theoretical calculation, square is the experiment [\cite{ilya}].}
\end{figure}

\nocite{*}
\bibliography{gutlian}

\end{document}